\documentclass[12pt]{article}
\usepackage{amsmath,amssymb}
\newtheorem{theorem}{Theorem}

\newtheorem{exe}[theorem]{Exercise}
\newtheorem{exa}[theorem]{Example}

\newtheorem{remark}[theorem]{Remark}

\newcommand{\CS}{{\cal S}}

\newcommand{\CN}{{\cal N}}

\newcommand{\CM}{{\cal M}}
\newcommand{\CQ}{{\cal Q}}

\def\al{\alpha}

\def\la{\lambda}

\newcommand{\rref}[1]{(\ref{#1})} 

\def\bih{bi-Ham\-il\-tonian}
\def\varb{\bih\ manifold}

\begin{document}

\title{On the bi-Hamiltonian structures\\ 
of the Camassa-Holm\\ and Harry Dym equations}
\author{Paolo Lorenzoni\\
Dipartimento di Matematica e Applicazioni\\
Universit\`a di Milano-Bicocca\\
Via Roberto Cozzi 53, I-20125 Milano, Italy\\
lorenzoni@matapp.unimib.it\\[2ex]
Marco Pedroni\\
Dipartimento di Matematica\\
Universit\`a di Genova\\   
Via Dodecaneso 35, I-16146 Genova, Italy\\
pedroni@dima.unige.it}

\maketitle

\begin{abstract}
We show that the bi-Hamiltonian structures of the Camassa-Holm and Harry Dym
 hierarchies can be obtained by applying a reduction
 process to a simple Poisson pair defined on the loop algebra
 of $\mathfrak{sl}(2,\mathbb{R})$.
 The reduction process is a bi-Hamiltonian reduction,
 that can be canonically performed on every bi-Hamiltonian manifold.  
\end{abstract}

\newpage
\section{Introduction}

In recent years a lot of papers have been devoted to the
Camassa-Holm equation (CH) 

\begin{equation}
u_{t}-u_{txx}=-3uu_{x}+2u_{x}u_{xx}+uu_{xxx}
\end{equation}
or, putting $m=u-u_{xx}$,
\begin{equation}
m_{t}=-2mu_{x}-m_{x}u,
\end{equation}
introduced in \cite{CH} as a model of shallow water waves.  
Part of them \cite{BSS1,BSS2,KM,McK}
 have investigated the connections with the
Korteweg-de Vries (KdV) equation
\begin{equation}
u_{t}=-3uu_{x}+u_{xxx},
\end{equation}
the Hunter-Saxton (HS) equation \cite{HS}
\begin{equation}
u_{txx}=-2u_{x}u_{xx}-uu_{xxx},
\end{equation}
and the Harry Dym (HD) equation
\begin{equation}
u_{t}=\left(u^{-\frac{1}{2}}\right)_{xxx},
\end{equation}
attributed in \cite{kruskal} to Harry Dym. 
In particular, Khesin and Misiolek \cite{KM}, motivated by \cite{BSS1} and \cite{BSS2},
 have explained the connections between KdV, CH with linear dispersion,
\begin{equation}
u_{t}-u_{txx}=-3uu_{x}+2u_{x}u_{xx}+uu_{xxx}+cu_{xxx},
\end{equation}
 and HS in terms of their common symmetry group, the Virasoro group.
 Indeed, these 
equations can be interpreted as Euler equations describing the geodesic 
flow (with respect to different metrics): on the Virasoro group
 in the KdV and CH ($c\ne 0$) case \cite{OK,Mi}, on the diffeomorphism group of the circle in the CH ($c=0$) case \cite{Mi,CK}, and on a suitable
 homogeneous space in the HS case \cite{KM}.
 Moreover, to any (codimension 2) coadjoint orbit there corresponds a
bi-Hamiltonian structure: the first Poisson bracket is just the
Lie-Poisson bracket, while the second one is a constant bracket depending
on the choice of a point in the dual of the Virasoro algebra. Points on 
the same orbit give rise to equivalent choices. There are three types of
orbit and three different associated bi-Hamiltonian structures: the KdV,
the CH and the HS bi-Hamiltonian structures. 

As regards the connections between CH and HD, they are clear in the 
framework of the inverse scattering techniques \cite{BSS1}. Indeed, both the 
equations are associated 
 to the scattering problems  for the family of operators
\begin{eqnarray*}
L_{k}=\partial_{x}^{2}+k^{2}\rho^{2}-q.
\end{eqnarray*}
A different choice of the boundary conditions for the function $\rho(x)$
 and of the value of the constant $q$ selects the associated equation: 
the case $q=0$ and $\rho\rightarrow 1$ at infinity corresponds to HD, while 
the case $q=\frac{1}{4}$ and $\rho\rightarrow 0$ at infinity corresponds 
to CH.

In this paper we investigate the connections between CH and HD
 from a different point of view.
More precisely, we show that the CH and HD bi-Hamiltonian structures
 can be obtained by a bi-Hamiltonian
reduction procedure from the Poisson pencil
\begin{eqnarray*}
P_{(\lambda)}=P_{2}+\lambda P_{1}=\partial_{x}+[\cdot,A]+\lambda[\cdot,S],
\end{eqnarray*}
defined on the space $\CM=C^\infty(S^1,\mathfrak{sl}(2,\mathbb{R}))$ of 
$C^{\infty}$ maps from the unit circle to the Lie algebra of $2\times 2$ 
traceless matrices.
 $S$ is a point of $\CM$ and $A$ is an arbitrary constant traceless 
matrix.
The reduction procedure depends on the choice of the conjugacy class of
$A$.

It turns out that there are three different reduced bi-Hamiltonian
structures: one is the CH bi-Hamiltonian structure ($\mbox{rank}(A)=2$), one is
the HD bi-Hamiltonian structure ($\mbox{rank}(A)=1$) and one, up to a change of
coordinates, is still the HD bi-Hamiltonian structure ($A=0$).

Since the HD bi-Hamiltonian structure can be obtained from HS 
bi-Hamiltonian structure just by a change of variables 
\cite{HZ}, and taking 
into account the correspondence between coadjoint Virasoro orbits and 
 Jordan normal forms in $SL(2,\mathbb{R})$, we
observe that this result seems to be strictly related to those of
Khesin and Misiolek.

The paper is organized as follows: In section 2 we summarize some useful techniques 
in the theory of the bi-Hamiltonian reduction. In section 3 we formalize and prove the above mentioned results.

\vspace{.5 cm}
\par\smallskip\noindent
{\bf Acknowledgments.}
We wish to thank Paolo Casati, Gregorio Falqui, and Franco Magri 
for useful discussions, and the referee for helpful remarks and corrections. 
 This work has been partially supported by INdAM-GNFM under the 
research project {\em Onde nonlineari, struttura tau e geometria 
delle variet\`a invarianti: il caso della gerarchia di Camassa-Holm.}

\section{The \bih\ reduction}

In this section we recall a reduction process of the Marsden-Ratiu type \cite{MR}, 
that can be performed on every \varb.
It has been presented in \cite{CMP} and then applied to the Drinfeld-Sokolov hierarchies \cite{DS} in \cite{CP,P} and to the stationary reductions of KdV in \cite{FMPZ}. 

Let $(\CM,P_1,P_2)$ be a \varb, i.e., a manifold $\CM$ endowed with two Poisson tensors $P_1$ and 
$P_2$ that are compatible, in the sense that their sum (and hence any linear combination) is still a Poisson tensor (see, e.g., \cite{cetraro,pondi}). Let us fix a symplectic leaf $\CS$ of $P_1$ and consider the distribution 
$D=P_2(\mbox{Ker}P_1)$ on $\CM$.

\begin{theorem} The distribution $D$ is integrable. If $E=D\cap T\CS$ is the distribution induced by $D$ on $\CS$ and the quotient space $\CN=\CS/E$ is a manifold, then it is a \varb.
\label{th1}
\end{theorem}

The reduced Poisson tensors $P_1^{\mbox{\scriptsize red}}$ and 
$P_2^{\mbox{\scriptsize red}}$ on $\CN$ are constructed as follows.
For any point $p\in\CS$ and any covector $\al\in T^*_{\pi(p)}\CN$, where $\pi:\CS\to\CN$ is the canonical projection, there is a covector $\tilde\al\in  T^*_{p}\CM$ such that 
\begin{equation*}
\tilde\al|_{D_p}=0\ ,\qquad \tilde\al|_{T_p\CS}=\pi^*_p\al\ ,
\end{equation*}
where $\pi^*_p:T^*_{\pi(p)}\CN\to T^*_{p}\CS$ is the codifferential of $\pi$ at $p$. Then
\begin{equation*}
\left(P_i^{\mbox{\scriptsize red}}\right)_{\pi(p)}\al={\pi_*}_p\left((P_i)_p\tilde\al\right)\ ,\qquad i=1,2\ .
\end{equation*}

Whenever an explicit description of the quotient manifold $\CN$ is not available,
 the following technique to compute the reduced \bih\ structure (already employed
 in \cite{CP} for the Drinfeld-Sokolov case) is very useful.

\begin{theorem}
 Suppose $\CQ$ to be a submanifold of $\CS$ which is transversal to the distribution $E$, in the sense that 
\begin{equation}
\label{split}
T_p\CQ\oplus E_p=T_p\CS\qquad\mbox{for all $p\in\CQ$}\ .
\end{equation}
Then $\CQ$ (which is locally diffeomorphic to $\CN$) also inherits a \bih\ structure from $\CM$. The reduced Poisson pair on $\CQ$ is given by 
\begin{equation}
\left(P_i^{\rm{rd}}\right)_{p}\al=\Pi_p\left((P_i)_p\tilde\al\right)\ ,\qquad i=1,2\ ,
\end{equation}
where $p\in\CQ$, $\al\in T^*_p\CQ$, $\Pi_p:T_p\CS\to T_p\CQ$ is the projection relative to \rref{split}, and $\tilde\al\in T^*_p\CM$ satisfies
\begin{equation}
\tilde\al|_{D_p}=0\ ,\qquad \tilde\al|_{T_p\CQ}=\al\ .
\end{equation}
\label{th2}
\end{theorem}

In the next section we will follow this procedure to construct the \bih\ structures of Camassa-Holm and Harry Dym as suitable \bih\ reduced structures.

\section{The \bih\ structure of CH and HD}

The aim of this section is to obtain the Poisson pair of Camassa-Holm and that of Harry Dym by applying the reduction procedure we have just described to a simple class of \bih\ structures on the loop algebra of $\mathfrak{sl}(2,\mathbb{R})$.

Let $\CM=C^\infty(S^1,\mathfrak{sl}(2,\mathbb{R}))$ be the space of $C^\infty$-maps from the unit circle to the Lie algebra of traceless $2\times 2$ real matrices. The tangent space $T_S\CM$ at $S\in\CM$ is obviously identified with 
$\CM$ itself. As far as the cotangent space is concerned, we will assume that $T^*_S\CM\simeq T_S\CM$ by means of the nondegenerate form
\begin{equation*}
\langle V_1,V_2\rangle=\int_{S^1}\mbox{tr}\left(V_1(x),V_2(x)\right)\,dx\ ,\qquad
V_1,V_2\in\CM\ .
\end{equation*}
As well-known (see, e.g., \cite{LM}), the manifold $\CM$ admits a 3-parameter family of compatible Poisson tensors given by 
\begin{equation}
V\mapsto \left(P_{(a,b,c)}\right)_S V=a\partial_x V+b[V,S]+c[V,A]\ ,\qquad
S\in\CM,\hspace{.5 cm} V\in T^*_S\CM\ ,
\end{equation}
where $a,b,c\in\mathbb{R}$ and $A$ is any matrix in $\mathfrak{sl}(2,\mathbb{R})$.
 We have the following theorems.

\begin{theorem}
 The \bih\ reduction process applied 
to the pair $(P_{1}=P_{(1,1,0)}, P_{2}=P_{(0,0,1)})$ gives rise:

- to the Poisson pair of the KdV hierarchy if $A=\begin{pmatrix} 0 & 0\\1 & 0\end{pmatrix}$;

- to the Poisson pair of the AKNS hierarchy if $A=\begin{pmatrix} 1 & 0\\0 & -1\end{pmatrix}$,

 for suitable choices of the symplectic leaf.
\end{theorem}
{\bf Proof}: See \cite{CMP} and \cite{MMR}. The reduction used in the 
latter paper is not the one presented in Theorem \ref{th1},
 but it is easily shown to be equivalent.

\begin{flushright}
$\Box$
\end{flushright} 

\begin{theorem}
The \bih\ reduction process applied to the pair $(P_1=P_{(0,1,0)},P_2=P_{(1,0,1)})$ 
gives rise:

- to the Poisson pair of the CH hierarchy if
 $A=\frac{1}{2}\begin{pmatrix} 0 & 1\\1 & 0\end{pmatrix}$;

- to the Poisson pair of the HD hierarchy if $A=\begin{pmatrix} 0 & 0\\1 & 0\end{pmatrix}$,

 for suitable choices of the symplectic leaf.
\end{theorem} 
{\bf Proof}: Let  
\begin{equation*}
S=\begin{pmatrix} p & q\\ r & -p\end{pmatrix}\ ,\qquad 
A=\begin{pmatrix} P & Q\\ R & -P\end{pmatrix}\ ,
\end{equation*}
with $p,q,r\in C^\infty(S^1,\mathbb{R})$ and $P,Q,R\in\mathbb{R}$. Since $\mbox{Ker}\left(P_1\right)_S$ is spanned 
by $S$, the symplectic leaves of $P_1$ are the level submanifolds of $\det S=-p^2-qr$. Moreover, we have that 
\begin{equation*}
D_S=(P_2)_S\left(\mbox{Ker}(P_1)_S\right)=\{(\mu S)_x+[\mu S,A]\mid \mu\in
C^\infty(S^1,\mathbb{R})\}\ .
\end{equation*}
Explicitly,
\begin{equation*}
D_S=\left\{\begin{pmatrix} (\mu p)_x+(Rq-Qr)\mu & (\mu q)_x+2(Qp-Pq)\mu \\ 
(\mu r)_x+2(Pr-Rp)\mu & -(\mu p)_x-(Rq-Qr)\mu \end{pmatrix}
\mid \mu\in C^\infty(S^1,\mathbb{R})\right\}\ .
\end{equation*}
The distribution $D$ is not tangent to the generic symplectic leaf of $P_1$,
 but it is easily shown to be 
tangent to the symplectic leaf 
\begin{equation}
\CS=\left\{\begin{pmatrix} p & q \\ r & -p \end{pmatrix}
\mid p^2+qr=0, (p,q,r)\not=(0,0,0)\right\}\ ,
\end{equation}
so that $E_p=D_p\cap T_p\CS$ coincides with $D_p$ for all $p\in\CS$. In order to determine the reduced \bih\ structure we first show that, under the assumption that $R\ne 0$, the submanifold
\begin{equation}
\CQ=\left\{\begin{pmatrix} 0 & q \\ 0 & 0 \end{pmatrix}
\mid q\in C^\infty(S^1,\mathbb{R}), q(x)\ne 0\ \forall x\in S^1\right\}
\end{equation}
of $\CS$ is transversal to the distribution $E$. Indeed, if $S(q)=\begin{pmatrix} 0 & q\\ 0 & 0\end{pmatrix}$, then
\begin{equation*}
T_{S(q)}\CS=\left\{\begin{pmatrix} \dot p & \dot q \\ 0 & -\dot p \end{pmatrix}
\mid \dot p,\dot q\in C^\infty(S^1,\mathbb{R})\right\}\simeq C^\infty(S^1,\mathbb{R})\oplus C^\infty(S^1,\mathbb{R})\ ,
\end{equation*}
and every tangent vector in $T_{S(q)}\CS$ admits the unique decomposition 
\begin{equation*}
(\dot p,\dot q)=(\dot p,\frac{1}{R}({\dot p}_x-2P\dot p))+(0,\dot q-\frac{1}{R}({\dot p}_x-2P\dot p))\ ,
\end{equation*}
where the first summand belongs to $E_{S(q)}$ and the second one to $T_{S(q)}\CQ$. This also shows that 
$\Pi_{S(q)}:T_{S(q)}\CS\to T_{S(q)}\CQ$ is given by
\begin{equation}
\Pi_{S(q)}:(\dot p,\dot q)\mapsto (0,\dot q-\frac{1}{R}({\dot p}_x-2P\dot p))\ .
\end{equation}
At this point we can compute the reduced Poisson pair on $\CQ$. For the sake of simplicity we will deal simultaneously with the Poisson pencil $P_{(\la)}=P_2+\la P_1$. Given $\al\in T^*_{S(q)}\CQ\simeq C^\infty(S^1,\mathbb{R})$, we look for a covector 
\begin{equation*}
\tilde\al=\begin{pmatrix} \al_1 & \al_2 \\ \al_3 & -\al_1 \end{pmatrix}\in T_{S(q)}^*\CM
\end{equation*}
such that $\tilde\al|_{D_{S(q)}}=0$ and $\tilde\al|_{T_{S(q)}\CQ}=\al$. We easily find that
\begin{equation*}
\tilde\al=\begin{pmatrix} \frac{1}{2R}(\al_x+2P\al) & \al_2 \\ \al & -\frac{1}{2R}(\al_x+2P\al) 
\end{pmatrix},
\end{equation*}
where $\al_2$ is arbitrary. Then we have that 
$\left(P_{(\la)}\right)_{S(q)}\tilde\al\in T_S\CQ$ is given by
\begin{eqnarray*}
&&\dot p=\frac{1}{2R}(\al_{xx}+2P\al_x)+R\al_2-Q\al-\la\al q\\
&&\dot q={\al_2}_x+\frac{Q}{R}({\al}_x+2P\al)-2P\al_2+\la\frac{q}{R}(\al_x+2P\al)\ .
\end{eqnarray*}
Thus the reduced Poisson pencil is
\begin{equation*}
\begin{split}
\left(P_{(\la)}^{\mbox{\scriptsize rd}}\right)_{q}\al 
&=\Pi_{S(q)}\left(\left(P_{(\la)}\right)_{S(q)}\tilde\al\right)\\
&=\left[-\frac{1}{{2R^2}}\partial_x^3+2\frac{QR+P^2}{R^2}\partial_x+
\frac{\la}{R}(2q\partial_x+q_x)\right]\al\ ,
\end{split}
\end{equation*}
that is to say
\begin{eqnarray*}
&&\left(P_1^{\mbox{\scriptsize rd}}\right)_{q} = \frac{1}{R}(2q\partial_x+q_x)\\
&&\left(P_2^{\mbox{\scriptsize rd}}\right)_{q} = -\frac{1}{2R^2}\partial_x^3
+2\frac{QR+P^2}{R^2}\partial_x \ .
\end{eqnarray*}

The case $P=0$, $Q=R=\frac{1}{2}$, i.e., $A=\frac{1}{2}\begin{pmatrix} 0 & 1\\ 1 & 
0\end{pmatrix}$,   corresponds to the Poisson pair 
\begin{eqnarray*}
&&\left(P_1^{\mbox{\scriptsize rd}}\right)_{q} =2(2q\partial_x+q_x)\\
&&\left(P_2^{\mbox{\scriptsize rd}}\right)_{q} =2( -\partial_x^3+\partial_x)\ . 
\end{eqnarray*}
It is well-known that it is the CH bi-Hamiltonian structure \cite{CH}.
Indeed, if we put $q=m=u-u_{xx}$:
\begin{equation}
m_{t}=-2mu_{x}-m_{x}u=P_1^{\mbox{\scriptsize rd}}\frac{\delta 
H_{1}}{\delta 
m}=P_2^{\mbox{\scriptsize rd}}\frac{\delta H_{2}}{\delta m},
\end{equation}
where 

\begin{eqnarray*}
&&H_{1}=-\frac{1}{4}\int(u^{2}+u^{2}_{x})dx,\\
&&H_{2}=-\frac{1}{4}\int(u^{3}+uu^{2}_{x})dx.
\end{eqnarray*}

The case $P=Q=0$, $R=1$, i.e., $A=\begin{pmatrix} 0 & 0\\ 1 & 0\end{pmatrix}$,
 gives rise to the Poisson pair
\begin{eqnarray*}
&&\left(P_1^{\mbox{\scriptsize rd}}\right)_{q} = 2q\partial_x+q_x\\
&&\left(P_2^{\mbox{\scriptsize rd}}\right)_{q} = -\frac{1}{2}\partial_x^3\ .
\end{eqnarray*}
It is well-known that it is the HD bi-Hamiltonian structure
 (see \cite{dorfman,magri,PSZ}).
Indeed, if we put $q=u$:
\begin{equation}
u_{t}=\left(u^{-\frac{1}{2}}\right)_{xxx}=P_1^{\mbox{\scriptsize 
rd}}\frac{\delta H_{1}}{\delta
u}=P_2^{\mbox{\scriptsize rd}}\frac{\delta H_{2}}{\delta u},
\end{equation}
where
\begin{eqnarray*}
&&H_{1}=\frac{1}{8}\int u^{-\frac{5}{2}}u_{x}^{2}dx,\\
&&H_{2}=-4\int u^{\frac{1}{2}}dx.
\end{eqnarray*}

\begin{flushright}
$\Box$
\end{flushright}

\begin{theorem}
The \bih\ reduction process applied
to the pair $(P_{1}=P_{(0,1,0)}, P_{2}=P_{(1,0,1)})$ 
with $A=\begin{pmatrix} 0 & 0\\0 & 0\end{pmatrix}$, in 
the space $ C^\infty_0(\mathbb{R},\mathfrak{sl}(2,\mathbb{R}))$
 of rapidly decreasing  $C^\infty$-maps from the real line 
to $\mathfrak{sl}(2,\mathbb{R})$, gives rise to the Poisson pair
\begin{equation}
(-2\left(\partial_{x}^{-1}p_{x}+p_{x} 
\partial_{x}^{-1}\right),\partial_{x}),
\end{equation}
where $p(x)$ is a smooth function vanishing at infinity and
\begin{equation*}
\partial_{x}^{-1}
=\frac{1}{2}\left(\int_{-\infty}^{x}-\int_{x}^{+\infty}\right).
\end{equation*}  
\end{theorem}
{\bf Proof}: The distribution
\begin{equation*}
D_S=\left\{\begin{pmatrix} (\mu p)_x  & (\mu q)_x \\
(\mu r)_x & -(\mu p)_x \end{pmatrix}
\mid \mu\in C^\infty_0(\mathbb{R},\mathbb{R})\right\}\ 
\end{equation*}
is still tangent to $\CS$. In this case the transversal submanifold is 
\begin{equation}
\CQ=\left\{\begin{pmatrix} p & 1 \\ -p^2 & -p \end{pmatrix}
\mid p\in C^\infty_0(\mathbb{R},\mathbb{R}), p(x)\ne 0\ \forall x\in 
\mathbb{R}\right\}
\end{equation}
and the projection $\Pi_{S(q)}:T_{S(q)}\CS\to T_{S(q)}\CQ$ is given by
\begin{equation}
\Pi_{S(q)}:(\dot p,\dot q)\mapsto 
(\dot{p}-p\dot{q}-p_{x}\partial_{x}^{-1}\dot{q},0).
\end{equation}
Following the same procedure used above it is easy to see that
\begin{eqnarray*}
&&\left(P_1^{\mbox{\scriptsize rd}}\right)_{p} = 
-2\left(\partial_{x}^{-1}p_{x}+p_{x}\partial_{x}^{-1}\right)\\
&&\left(P_2^{\mbox{\scriptsize rd}}\right)_{p} = 
\partial_{x}.
\end{eqnarray*} 
Moreover, taking into account that,
 after a change of coordinates $u'=u'(u)$, a 
bivector $P$ transforms as
\begin{equation}
P'=\left(\sum_{s\ge 0}\frac{\partial u'}{\partial 
u^{(s)}}\partial_{x}^{s}\right)
P\left(\sum_{t\ge 0}(-\partial_{x})^{t}\frac{\partial u'}{\partial 
u^{(t)}}\right),
\end{equation}
we obtain that, in the variable $u=p_{x}$, the Poisson pair 
$(P_1^{\mbox{\scriptsize rd}},P_2^{\mbox{\scriptsize rd}})$ coincides 
with the Harry Dym bi-Hamiltonian structure.

\begin{flushright}
$\Box$
\end{flushright}

\section{Conclusions}

In this paper we have shown that the bi-Hamiltonian structures of
 CH and HD can be seen as reductions of suitable structures on 
$C^\infty(S^1,\mathfrak{sl}(2,\mathbb{R}))$. In the KdV case this is 
a well-known result, that can be interpreted both from the Drinfeld-Sokolov 
point of view and in the bi-Hamiltonian reduction scheme \cite{P}.

The results of this paper could be used to construct 2-field extensions of the CH and HD hierarchies, in the same way as Drinfeld and 
Sokolov construct a 2-component generalization of the KdV hierarchy. This extended hierarchy lives on a symplectic leaf of 
$P_{(0,0,1)}$ and 
projects on the usual scalar KdV hierarchy (see \cite{DS} and \cite{pondi}). Analogously, we plan to define a 2-field hierarchy on the 
symplectic leaf $\CS$ of CH, whose projection on the transversal submanifold $\CQ$ is the CH hierarchy. The ``unprojected'' CH equation 
should be compared with the 2-component generalization of CH recently introduced by Liu and Zhang \cite{LZ}. 

A crucial point in Theorem 4 is the choice of the symplectic leaf $\CS$. As we have already said, this leaf is special in that the distribution $D$ is tangent to it. Thus, for a different leaf the reduction process is more complicated and turns out to be similar to the AKNS case. It would be interesting to construct the equations corresponding to these alternative choices.


\begin{thebibliography}{10}

\footnotesize

\bibitem{BSS1} R. Beals, D. Sattinger and J. Szmigielski,
 {\em Acoustic scattering and the extended Koteweg de Vries hierarchy\/}, 
Adv. in Math., {\bf 140} (1998), 190-206.

\bibitem{BSS2} R. Beals, D. Sattinger and J. Szmigielski,
{\em Inverse scattering solutions of the Hunter-Saxton equations\/},
 Appl. Anal. {\bf 78}  (2001), 255-269.

\bibitem{CH}
R. Camassa and D. Holm,
{\em An integrable shallow water equation with peaked solitons\/}, 
Phys. Lett. Rev. {\bf 71} (1993), 1661-1664.

\bibitem{CMP}
P. Casati, F. Magri, M. Pedroni,
{\em Bi-Hamiltonian Manifolds and $\tau$--function\/},
in: Mathematical Aspects of Classical
Field Theory 1991 (M. J. Gotay et al.\ eds.),
Contemporary Mathematics vol. {\bf 132},
American Mathematical Society,
Providence, R.I., 1992, pp.\ 213--234.

\bibitem{CP}
P. Casati and M. Pedroni,
{\em Drinfeld--Sokolov Reduction on a
Simple Lie Algebra from the Bi-Hamiltonian Point of View\/},
Lett. Math. Phys. {\bf 25} (1992), 89-101.

\bibitem{CK}
A. Constantin and B. Kolev, 
{\em On the geometric approach to the motion of inertial mechanical systems\/},
 J. Phys. A: Math. Gen. {\bf 35} (2002), R51-R79.

\bibitem{dorfman} I. Dorfman, {\em Dirac structures and integrability of
 nonlinear evolution equations\/}, John Wiley and Sons Ltd,
Chichester, 1993.

\bibitem{DS}
V. G. Drinfeld, V. V. Sokolov,
{\em Lie Algebras and Equations of Korteweg--de Vries Type\/},
J. Sov. Math. {\bf 30} (1985), 1975--2036.

\bibitem{FMPZ} G. Falqui, F. Magri, M. Pedroni, J.P. Zubelli,
{\em A Bi-Hamiltonian Theory for Stationary KdV Flows and their Separability\/},
Regul.\ Chaotic Dyn.\ {\bf 5} (2000), 33-52.

\bibitem{HS} J. Hunter and R. Saxton, {\em Dynamics of director fields\/}, 
SIAM J. Appl. Math. {\bf 51} (1991), 1498-1521.

\bibitem{HZ} J. Hunter and Y. Zheng, {\em On a completely integrable 
nonlinear variational equation\/}, Phys. D\ {\bf 79} (1994) 361-386

\bibitem{KM} B. Khesin and G. Misiolek, {\em Euler equations on homogeneous 
spaces and Virasoro orbits\/},  Adv.\ Math.\  {\bf 176}  (2003), 116-144. 

\bibitem{kruskal} M. Kruskal, {\em Nonlinear wave equations\/},
 Dynamical Systems, Theory and Applications (J. Moser
ed.), Lecture Notes in Phys.  
{\bf 38}, Springer, Heidelberg, 1975, pp. 310-354.  

\bibitem{LM} P. Libermann and C. M. Marle, {\em Symplectic Geometry and
 Analytical Mechanics\/}, Reidel, Dordrecht, 1987.

\bibitem{LZ} S. Liu and Y. Zhang, {\em Deformations of semisimple bi-Hamiltonian
 structures of hydrodynamic type}, math.DG/0405146.

\bibitem{magri} F. Magri, {\em A geometrical approach to the nonlinear 
solvable equations\/},  Nonlinear evolution equations and dynamical systems
 (Proc. Meeting, Univ. Lecce, Lecce, 1979), 
Lecture Notes in Phys. {\bf 120}, 1980, pp. 233-263.

\bibitem{pondi}
F. Magri, P. Casati, G. Falqui and M. Pedroni, 
{\em Eight lectures on Integrable Systems\/},
in: Integrability of Nonlinear Systems (Y. Kosmann-Schwarzbach et
al. eds.), Lecture Notes in Physics {\bf 495} (2nd edition),
 2004, pp.\ 209--250.

\bibitem{cetraro} 
F.\ Magri, G.\ Falqui, M.\ Pedroni, 
{\em The method of Poisson pairs in the theory of nonlinear 
PDEs\/}, in: Direct and Inverse Methods in Nonlinear Evolution Equations,
Lectures Given at the C.I.M.E. Summer School Held in Cetraro, Italy, 1999,
Lecture Notes in Physics {\bf 632}, 2003; nlin.SI/0002009.

\bibitem{MMR}
F. Magri, C. Morosi and O. Ragnisco, {\em Reduction techniques for infinite-
dimensional Hamiltonian systems: some ideas and applications\/},
 Commun. Math. Phys. {\bf 99} (1985), 115-140.

\bibitem{MR}
J.E. Marsden, T. Ratiu,
{\em Reduction of Poisson Manifolds\/},
Lett. Math. Phys. {\bf 11} (1986), 161--169.

\bibitem{McK}
H. McKean, 
{\em The Liouville correspondence between the Korteweg-de Vries and
 the Camassa-Holm hierarchies\/},
 Comm. Pure Appl. Math.  {\bf 56}  (2003), 998-1015. 

\bibitem{Mi} G. Misiolek, {\em A shallow water equation as a geodesic
 flow on the Bott-Virasoro group\/}, J. Geom. Phys. {\bf 24} (1998), 203-208.

\bibitem{OK} V. Ovsienko and B. Khesin, {\em The (super) KdV equation
 as an Euler equation\/}, Funct. Anal. Appl. {\bf 21} (1987), 81-82.

\bibitem{P} 
M. Pedroni
{\em Equivalence of the Drinfeld-Sokolov reduction to a bihamiltonian
reduction\/},
Lett. Math. Phys. {\bf 35} (1995), 291-302.

\bibitem{PSZ} 
M. Pedroni, V. Sciacca, J.P. Zubelli, 
{\em On the Bi-Hamiltonian Theory for the Harry Dym Equation\/}, Theor. Math. Phys.
 {\bf 133} (2002), 1583-1595.

\end{thebibliography}
\end{document}